\documentclass[english]{article}
\usepackage[T1]{fontenc}
\usepackage[latin9]{inputenc}
\usepackage{graphicx}
\usepackage{babel}
\begin{document}

\title{Mirror bootstrap method for testing hypotheses of one mean }

\author{Anna Varvak%
\thanks{Soka University of America, Aliso Viejo CA, USA, avarvak@soka.edu%
}}
\maketitle
\begin{abstract}
The general philosophy for bootstrap or permutation methods for testing
hypotheses is to simulate the variation of the test statistic by generating
the sampling distribution which assumes both that the null hypothesis
is true, and that the data in the sample is somehow representative
of the population. This philosophy is inapplicable for testing hypotheses
for a single parameter like the population mean, since the two assumptions
are contradictory (e.g., how can we assume both that the mean of the
population is $\mu_{0},$ and that the individuals in the sample with
a mean $M\ne\mu_{0}$ are representative of the population?). The
Mirror Bootstrap resolves that conundrum. The philosophy of the Mirror
Bootstrap method for testing hypotheses regarding one population parameter
is that we assume both that the null hypothesis is true, and that
the individuals in our sample are as representative as they could
be without assuming more extreme cases than observed. For example,
the Mirror Bootstrap method for testing hypotheses of one mean uses
a generated symmetric distribution constructed by reflecting the original
sample around the hypothesized population mean $\mu_{0}$. Simulations
of the performance of the Mirror Bootstrap for testing hypotheses
of one mean show that, while the method is slightly on the conservative
side for very small samples, its validity and power quickly approach
that of the widely used t-test. The philosophy of the Mirror Bootstrap
is sufficiently general to be adapted for testing hypotheses about
other parameters; this exploration is left for future research.
\end{abstract}

\section{Mirror Bootstrap method}

The general philosophy for bootstrap or permutation methods for testing
hypotheses is to simulate the variation of the test statistic by generating
the sampling distribution which assumes both that the null hypothesis
is true, and that the data in the sample is somehow representative
of the population. This philosophy works well for testing hypotheses
regarding the correlation coefficient, but is inapplicable for testing
hypotheses for a single parameter like the population mean, since
the two assumptions are contradictory. For example, how can we assume
both that the mean of the population is $\mu_{0},$ and that the individuals
in the sample with a mean $M\ne\mu_{0}$ are representative of the
population? One naive way that has been used is the Shift method \cite{Efron 1994},
where each individual in the sample is shifted by $\mu_{0}-M$, which
is essentially equivalent to testing the hypotheses with a confidence
interval. The Shift method hypothesizes that the variance between
the sampled individuals is representative of the population variance,
yet it loses any semblance to the assumption that the sampled individuals
themselves are representative of the population.

The Bootstrap Method provides a more elegant resolution. The philosophy
of the Mirror Bootstrap method for testing hypotheses regarding one
population parameter is to assume both that the null hypothesis is
true, and that the individuals in our sample are as representative
as they could be without assuming more extreme cases than observed.

For example, the Mirror Bootstrap method for testing hypotheses of
one mean uses a generated symmetric distribution constructed by reflecting
the original sample around the hypothesized population mean $\mu_{0}$.
The rationale for the Mirror Bootstrap is that, on the one hand, to
carry out a test against the null hypothesis one assumes that it's
true and computes the rarity of the evidence, and on the other hand
bootstrap methods try to assume that the sample is in some ways representational of the population. Obviously for a test of one mean, one can't assume both that the true population mean is $\mu_{0}$ and that the sampled
individuals are representational of the population (except in the
rare cases where the sample mean equals $\mu_{0}$, and those cases
are not interesting). But we can do the next best thing: assume that
the individuals sampled are as representational as they can be with
the null hypothesis being true. If we also don't want to assume individuals
more extreme than those observed, the easiest way to do that is to
assume that for every observed individual there is a partner somewhere
in the population who is on the opposite side of the hypothesized
population mean $\mu_{0}$. Thus, suppose we have a random sample
of size $n$ and sample mean $M$ of a random variable $X$ whose
population distribution is unknown, and we wish to test against the
null hypothesis $H_{0}:\mu=\mu_{0}$ with some alpha level. The Mirror
Bootstrap method goes as follows: reflect the sample around $x=\mu_{0}$,
getting a symmetric sample of size $2n$; then repeatedly sample with
replacement samples of size $n$ from this symmetric constructed population,
counting how many times the means of these bootstrap samples are further
from $\mu_{0}$ than the original sample mean $M$; if this proportion
is less than alpha, reject the null hypothesis.

How well does the Bootstrap Method work in practice? Simulations of
the performance of the Mirror Bootstrap for testing hypotheses of
one mean show that, while the method is slightly on the conservative
side for very small samples, its validity and power quickly approach
that of the widely used t-test. The author used Maple 15 to implement
the simulations, and would gladly share the scripts upon demand.

\section{Test of power and validity in three simple cases}

For a quick check of the power and validity of the Mirror Bootstrap
method, we compare its performance to that of the widely used two-tailed
t-test and to the Shift bootstrap test, in the cases where the population
is normally distributed, very skewed, and bimodal. The results indicate
that the Shift bootstrap test has serious problems with validity for
small samples ($n\le20)$; the Mirror Bootstrap method is more conservative
than the t-test for very small samples ($n\le5$) but matches the
validity and power of the t-test for samples size $n\ge10$. To represent
these three cases, we use the following distributions: standard normal
distribution $N(0,1)$, highly skewed distribution Gamma(2,2), and
a bimodal distribution $N(-3,1)$ XOR $N(3,1)$, all normalized so
that the population means are at 0, and the population standard deviations
are 1.

\includegraphics[width=11cm]{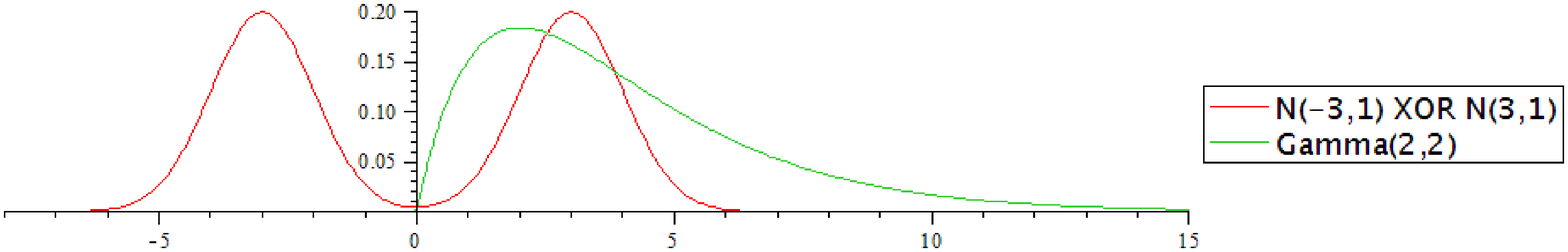}

\subsection{Tests for validity}

We simulated drawing a sample of specified size 10,000 times in each
case, bootstrapping 1000 samples, and doing two-tailed hypothesis
tests with alpha level 0.05. The results are as follows.

\includegraphics[width=6cm]{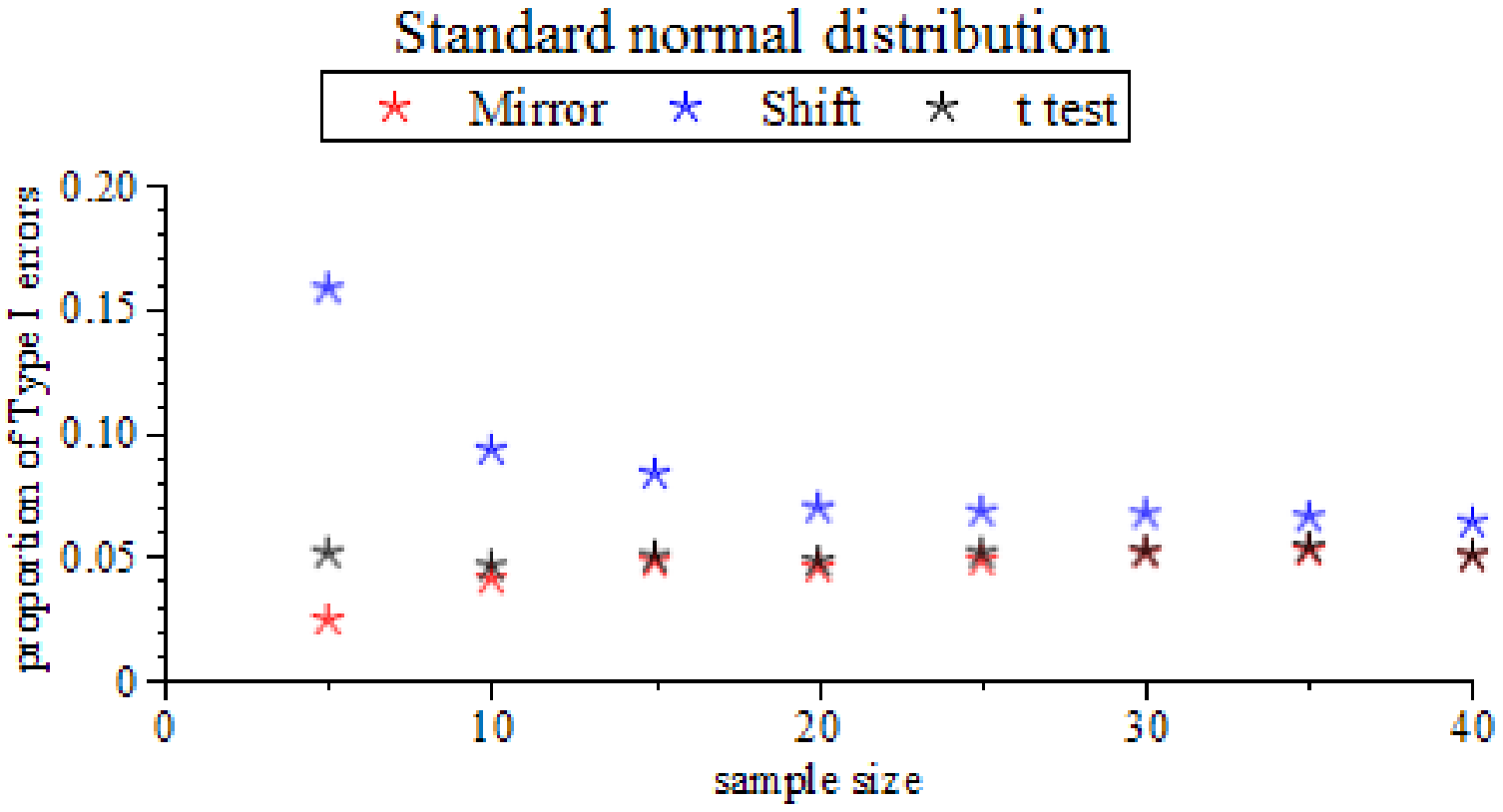}
\includegraphics[width=6cm]{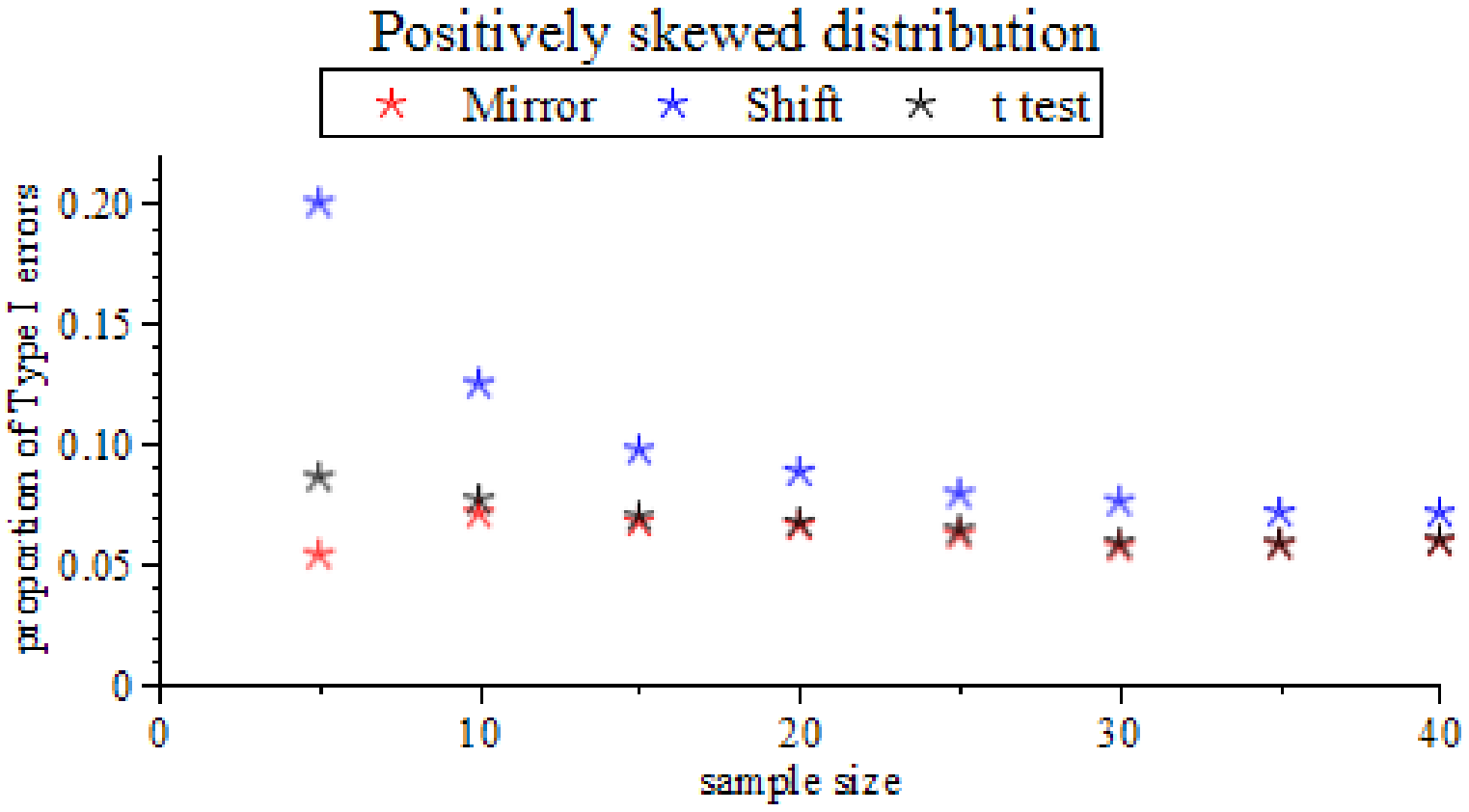}

\includegraphics[width=6cm]{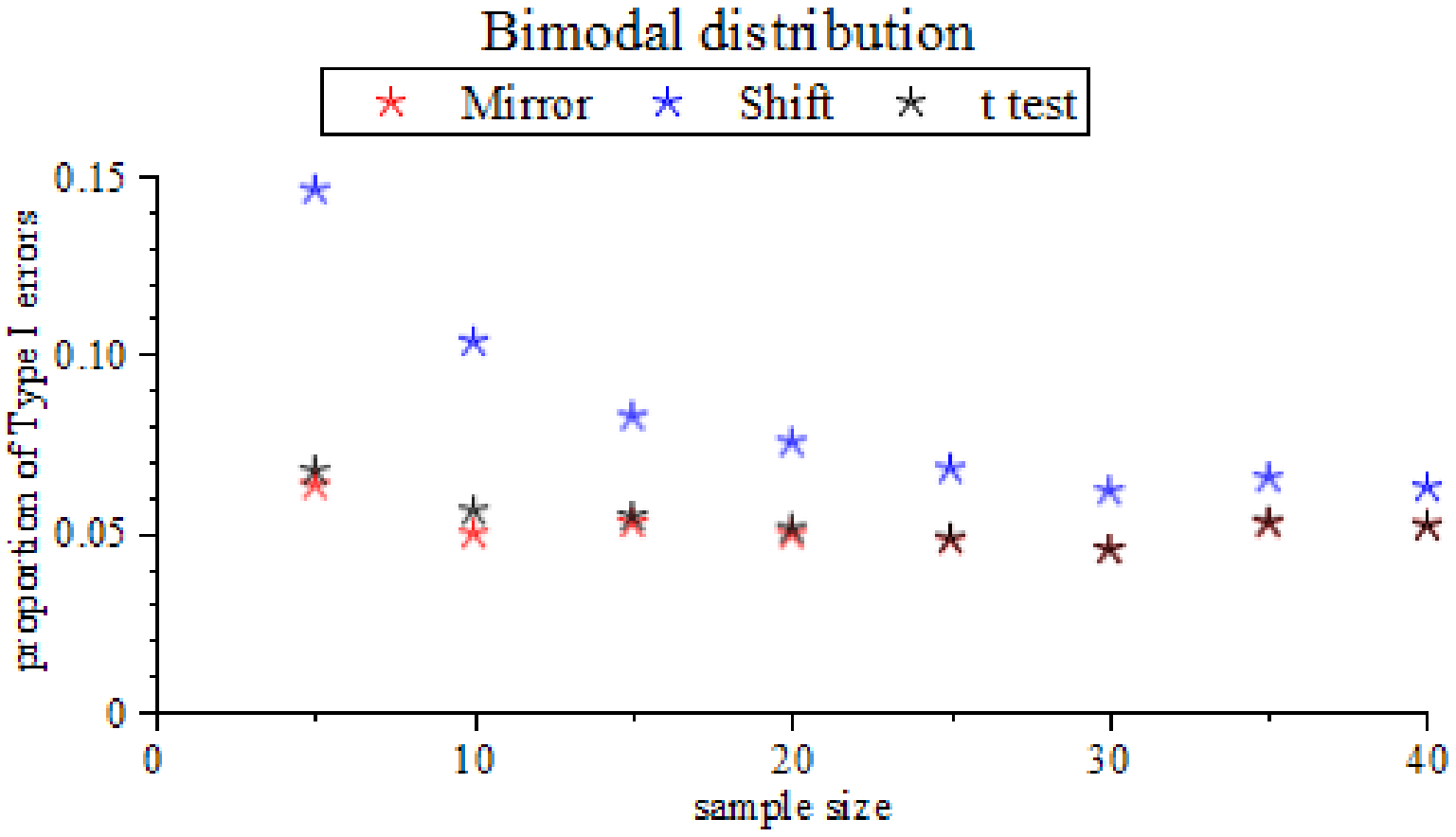}

For normal and skewed populations, mirror bootstrap is conservative
for very small samples, after which it performs comparably or better
than the t-test. It shouldn't come as a surprise that the performance
of the t-test is unbeaten for the normal distribution, since the t-test
was specifically designed with the assumption of normality. For a
bimodal population, mirror bootstrap isn't as conservative, and slightly
overshoots the target alpha of 0.05 for small samples of size 5. It
performs comparably or better than the t-test.

\subsection{Tests for power}

We simulated drawing a sample of specified size 1000 times in each
case, bootstrapping 1000 samples, and doing two-tailed hypothesis
tests with alpha level of 0.05. The results are as follows.

\includegraphics[width=6cm,height=6cm]{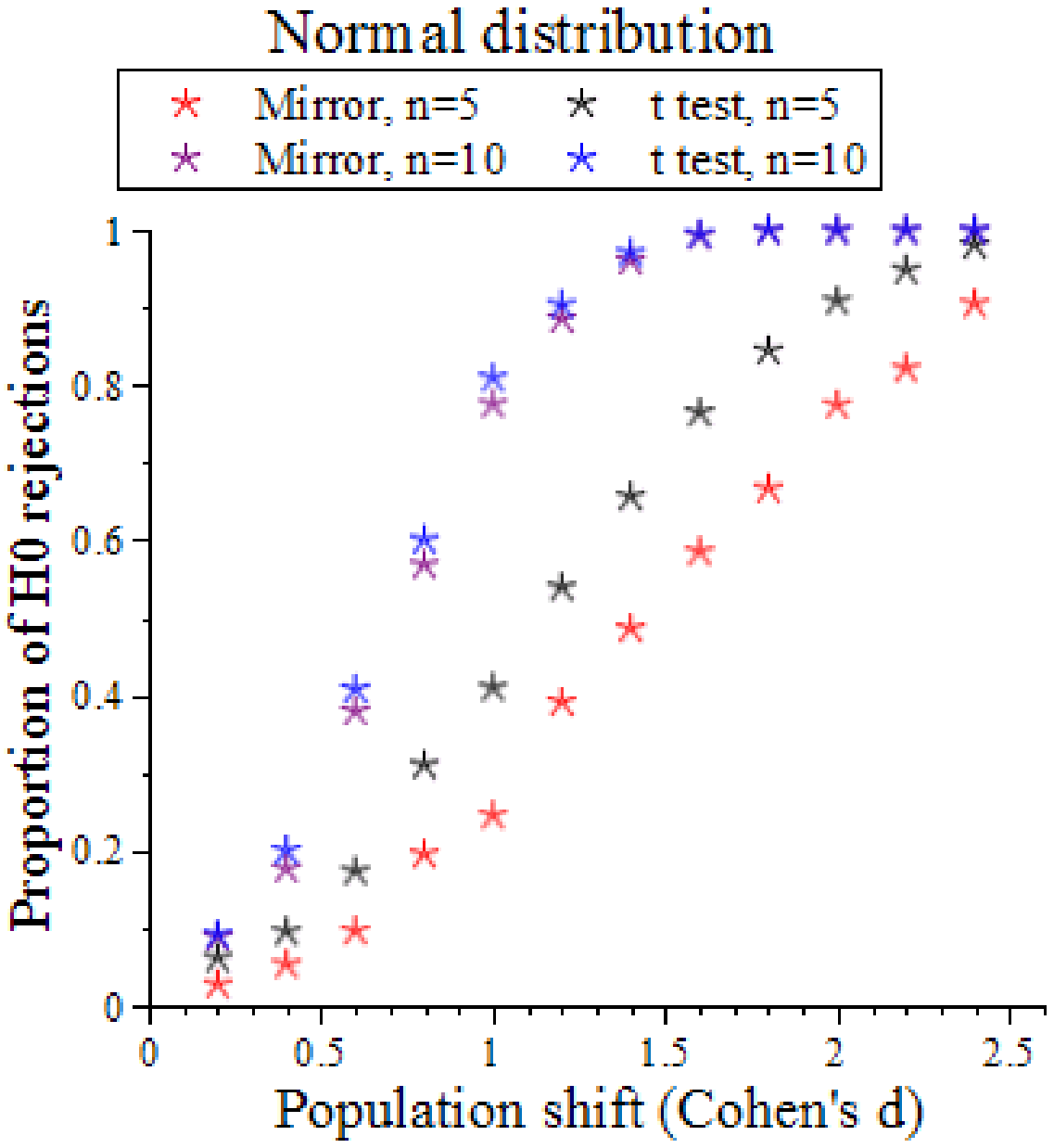}
\includegraphics[width=6cm,height=6cm]{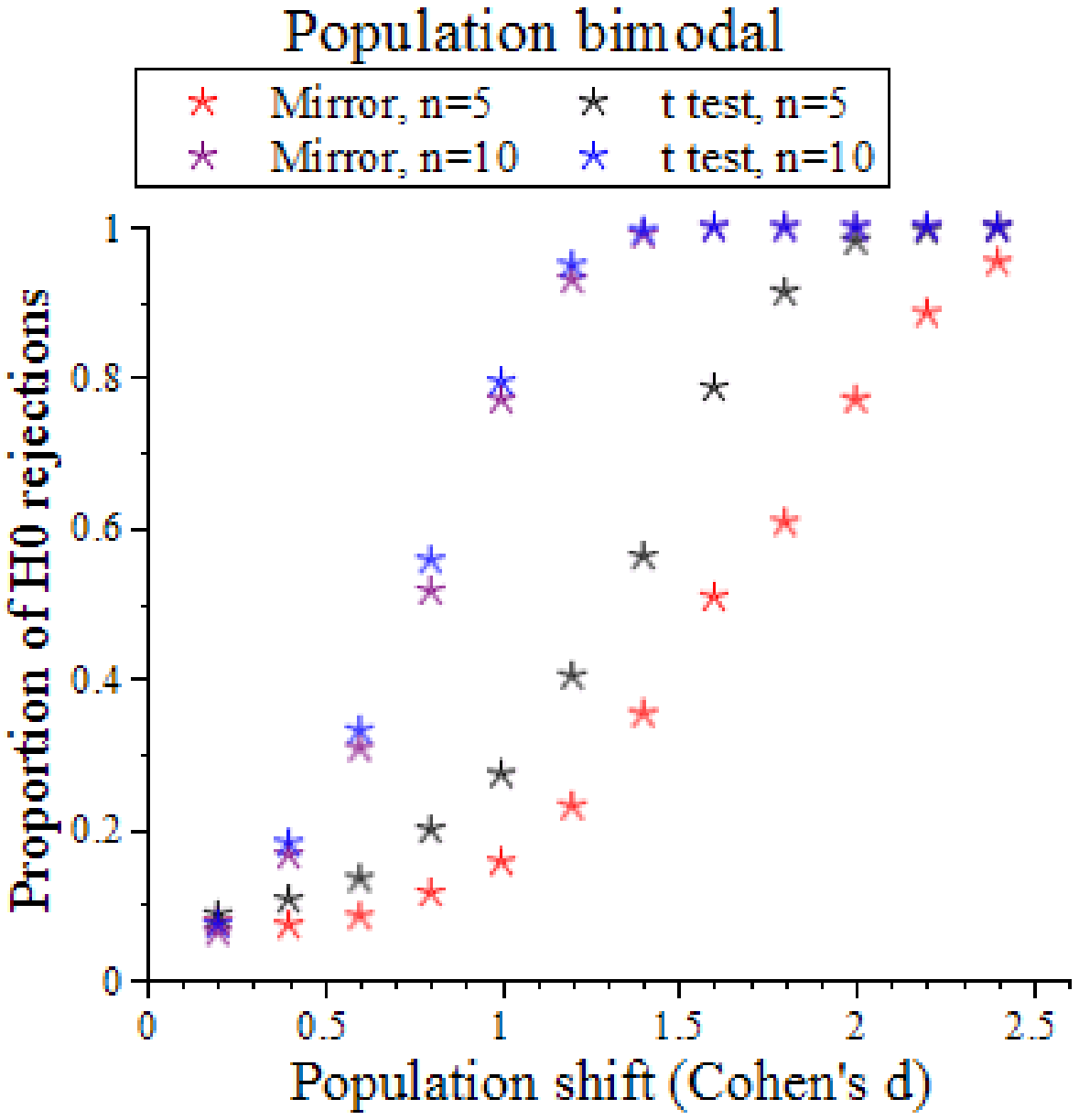}

\includegraphics[width=6cm,height=6cm]{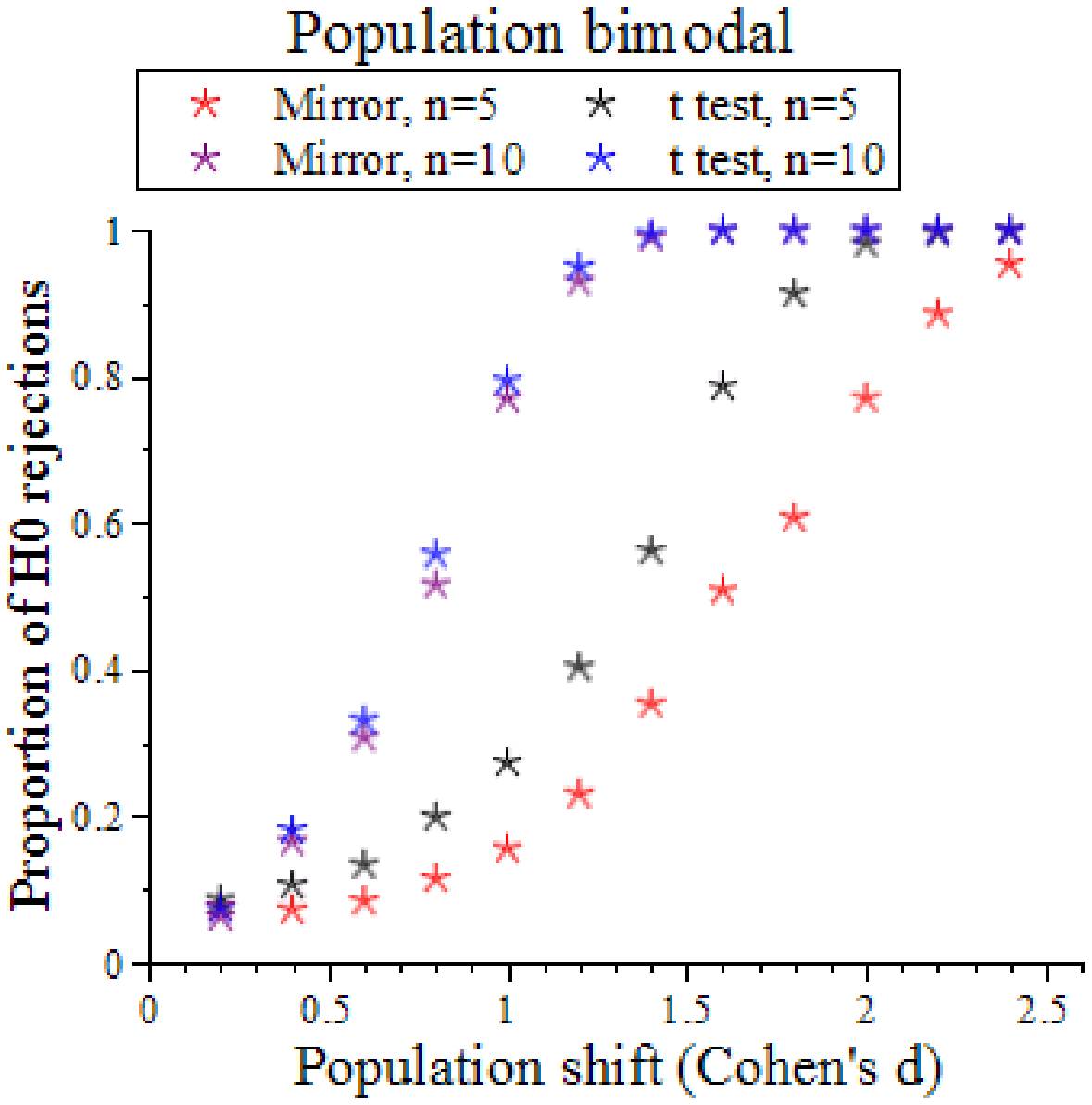}

For the distribution skewed to the left, we used a mirror version
of the normalized Gamma(2,2) distribution. For all three distributions,
the power curve for Mirror Bootstrap is slower for small samples of
size five, but are seen to converge to the power curve of the t-test
for samples of size 10.

An interesting result happens when looking at the case where the population
is heavily skewed to the right, as is the case for the normalized
Gamma(2,2):

\includegraphics[width=6cm,height=6cm]{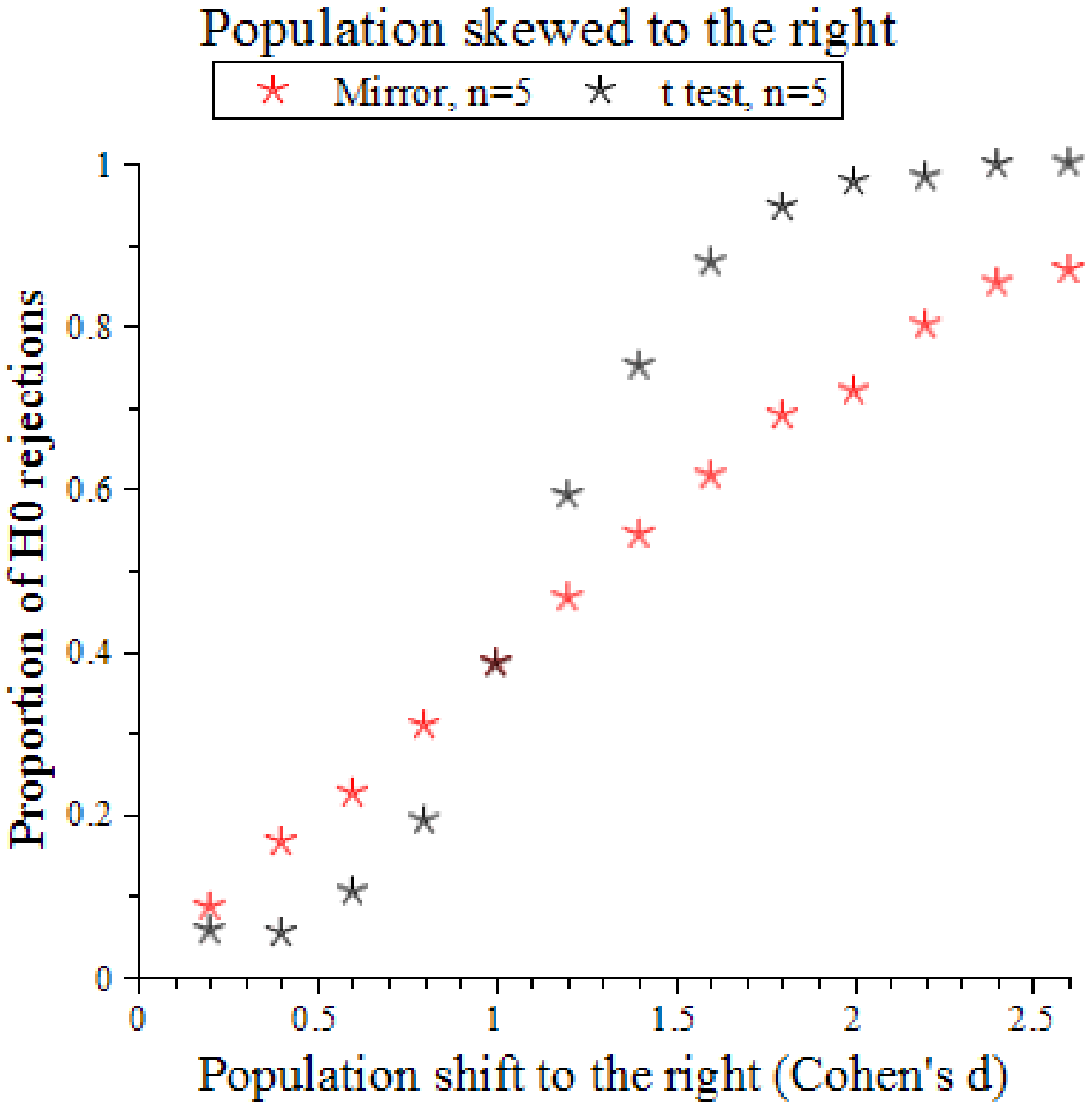}
\includegraphics[width=6cm,height=6cm]{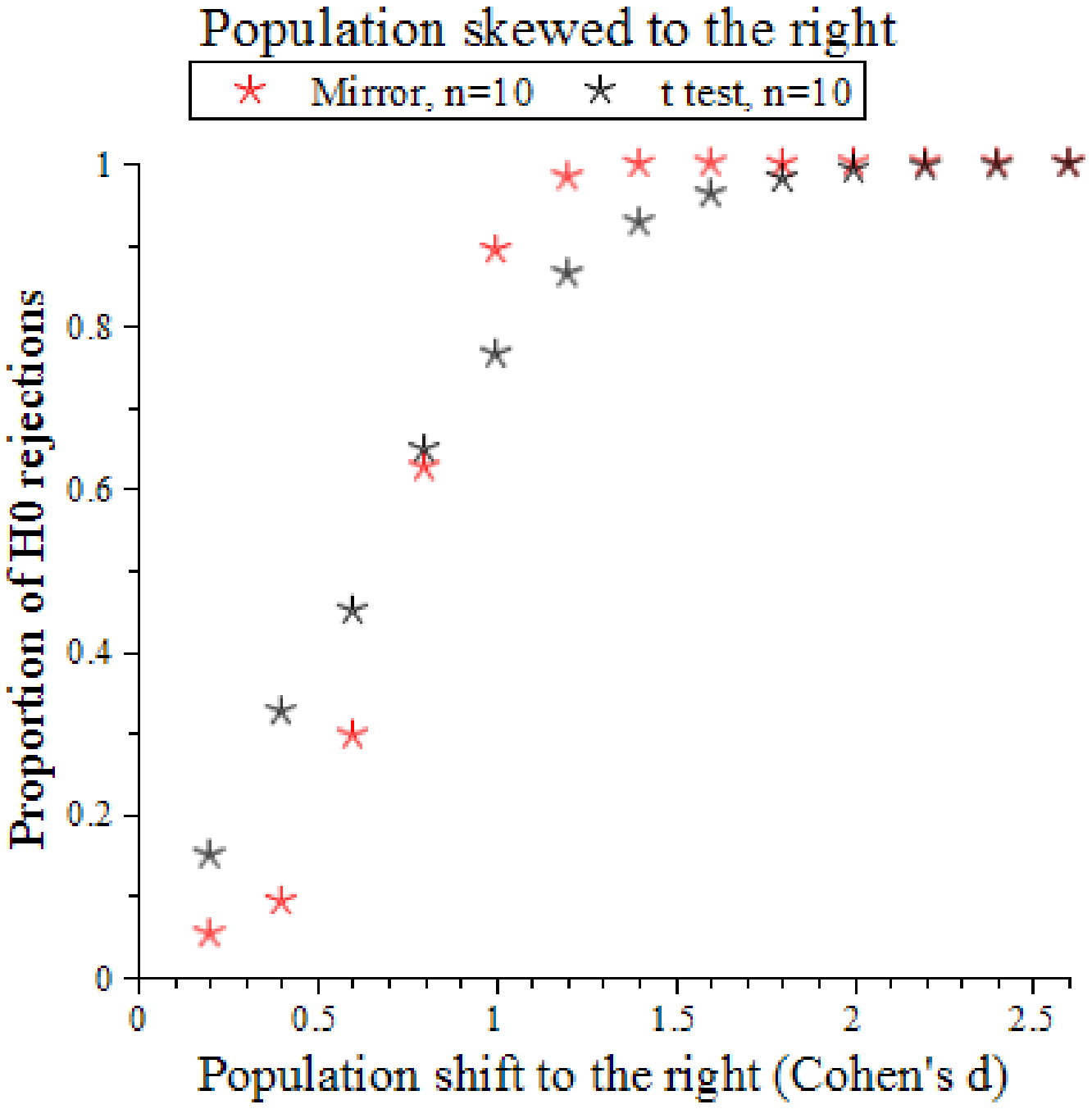}

In some situations, the Mirror Bootstrap outperforms the t-test.

\section{Systematic assessment of validity, using g-and-h distributions}

For a more systematic assessment of the Mirror Bootstrap method's
validity and power, we turn to the family of g-and-h distributions:
\[
X=\frac{\exp(gZ)-1}{g}\exp\left(\frac{hZ^{2}}{2}\right),
\]
which, in the case of $g=0,$ is $X=Z\exp\left(hZ^{2}/2\right)$.
Depending on the parameters $g$ and $h$, the distributions vary
in skewness and kurtosis. The family includes the standard normal
distribution, the lognormal distribution with its long skinny tail,
a symmetric distribution with heavy tails, a skewed symmetric distribution
with heavy tails, and everything in-between \cite{Hoaglin 1985,Wilcox 2012}.
Skewness is measured by $\mu_{3}/\sqrt{(\mu_{2}^{3})}$ and kurtosis
is measured by $\mu_{4}/\mu_{2}^{2}$, where the $\mu_{i}$ are the
moments around the mean, defined as usual: $\mu=E[X]$, and $\mu_{k}=E[(X-\mu)^{k}]$.
Following \cite{Hoaglin 1985}, the skewness and kurtosis of various
g-and-h distributions are as follows, with skewness of 0 whenever
$g=0$ and undefined for $h\ge1/3$, and kurtosis undefined for $h\ge1/4$:

\includegraphics[width=10cm]{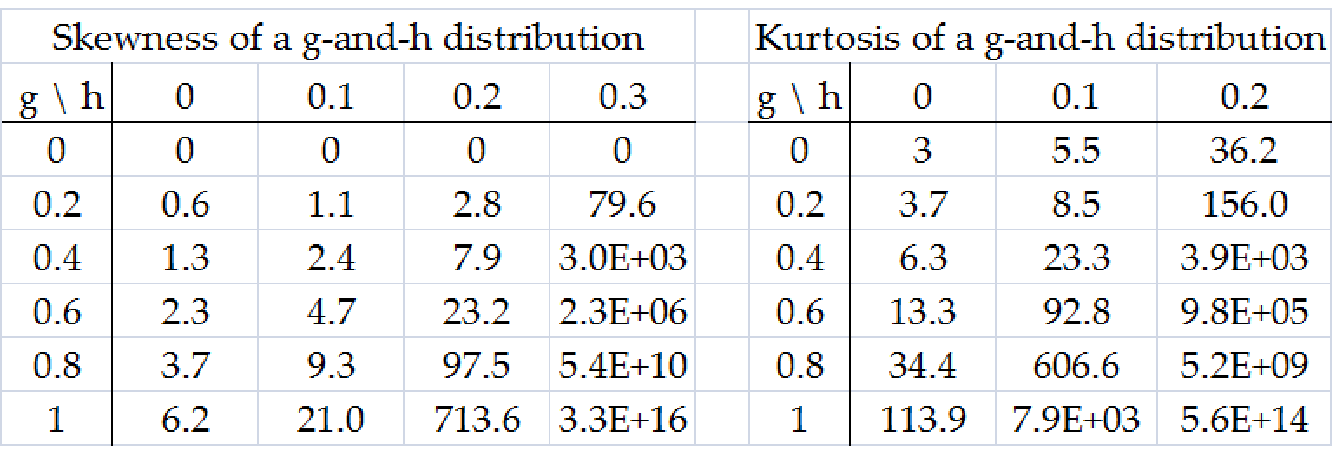}

\subsection{A useful case: $g=h=0.5$}

The g-and-h distribution with $g=h=0.5$ is an example of a very skewed
distribution with heavy tail. The skewness and kurtosis are undefined
for the population with such distribution; the mean is approximately
0.8. Wilcox \cite{Wilcox 2012} notes that a sample drawn from such
population has statistics that do not correspond to the population
parameters, showing in particular that a sample of 100,000 observations
had skewness of 120 and kurtosis of about 18,400. For this distribution,
power isn't the issue for the Mirror Bootstrap or the t-test. Power
curves are comparable, even for small samples.

\includegraphics[width=6cm,height=6cm]{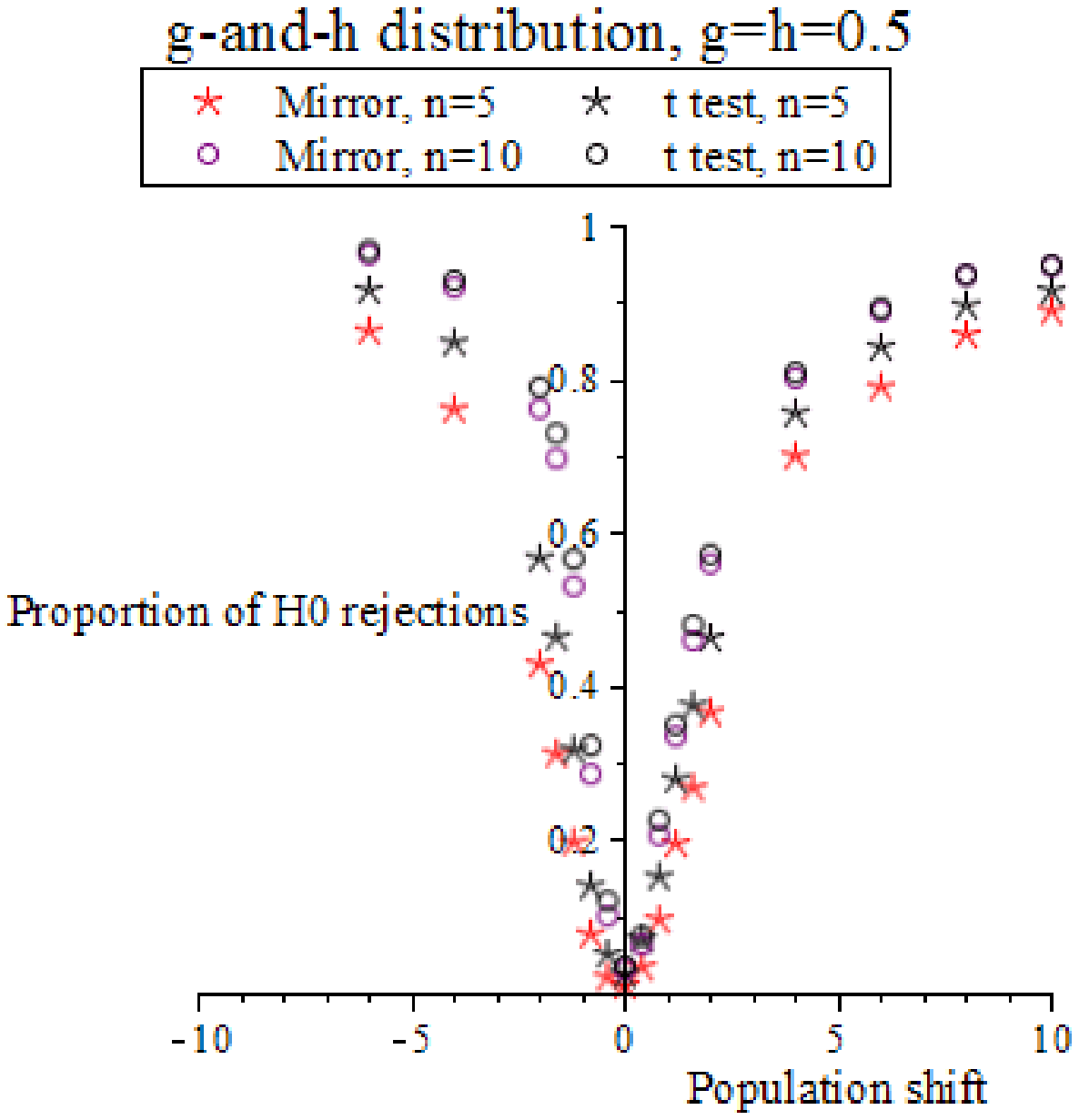}
\includegraphics[width=6cm,height=6cm]{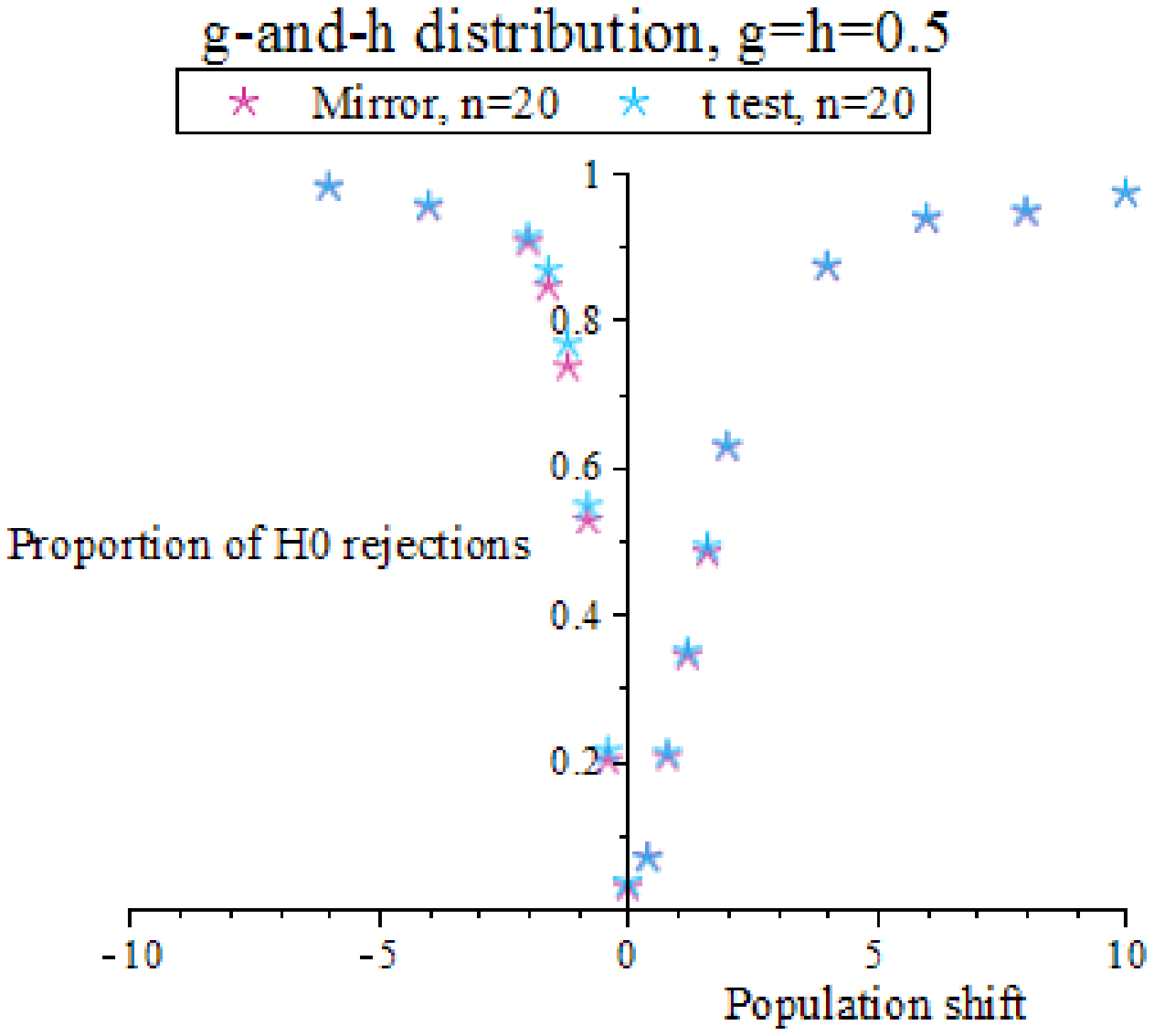}

Validity is a real problem for both methods: Mirror bootstrap performs
a little better than the t-test for small samples, though still performing
much higher than the assumed alpha level of 0.05. For large samples,
both methods plateau with approximately 20\% of Type I errors for
samples of size less than 100, and very slowly descend as the sample
size grows.

\includegraphics[width=5cm,height=5cm]{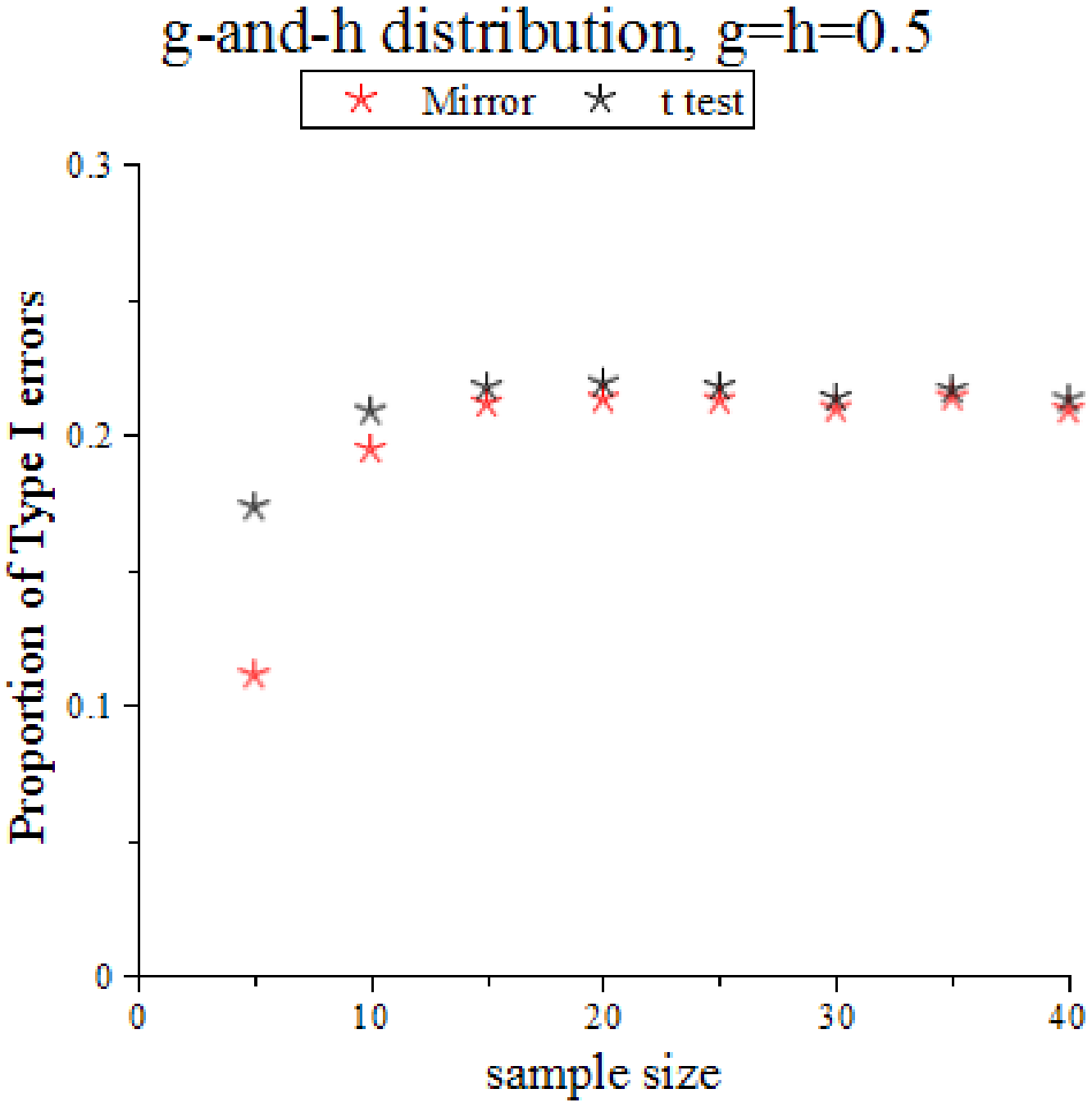}
\includegraphics[width=5cm,height=5cm]{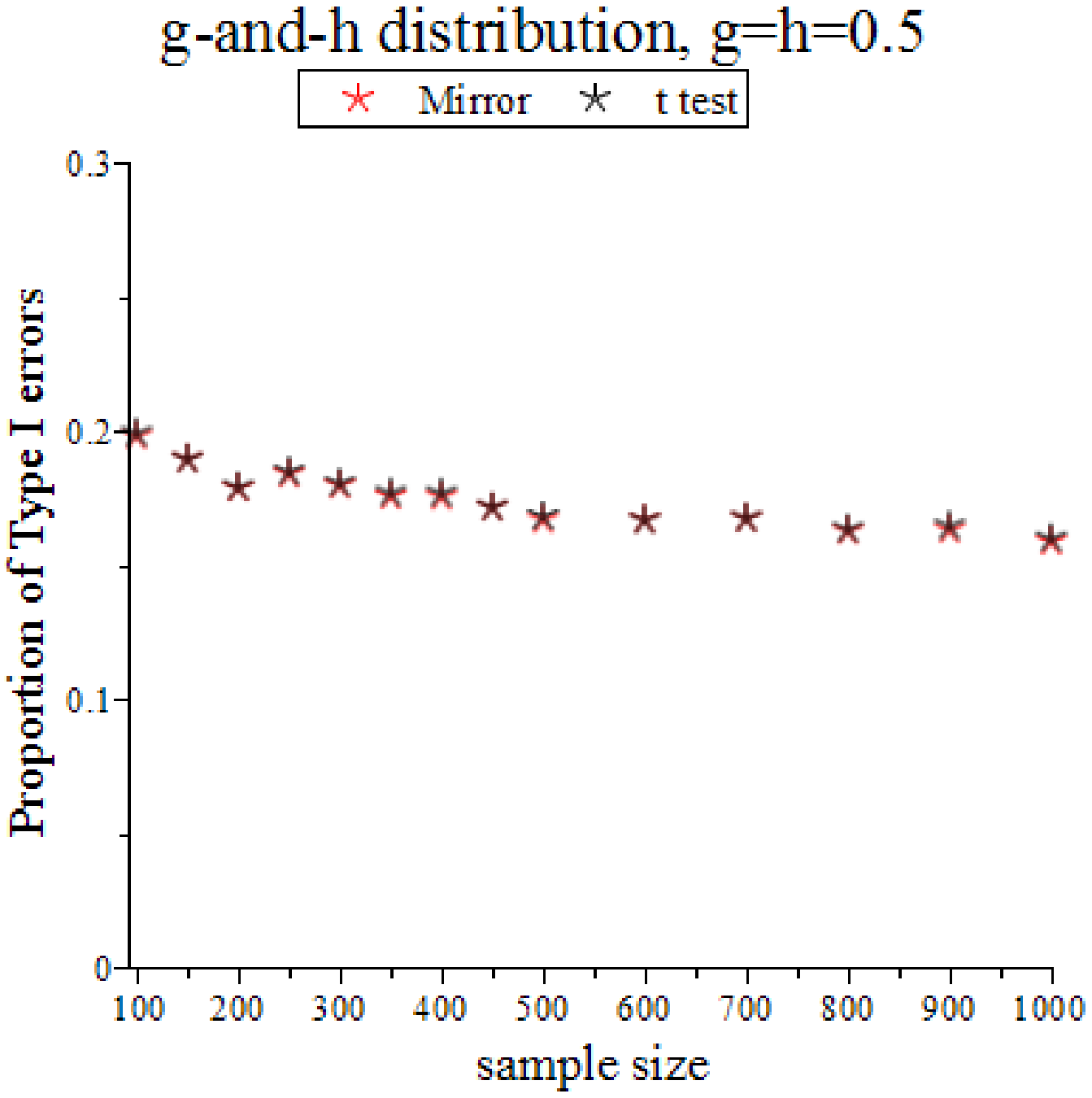}

In practice, one may ask if the mean is even an appropriate measure
of central tendency for such a distribution.

\subsection{Skewness without heavy tails ($h=0$)}

To test the validity of the Mirror Bootstrap method in testing hypotheses
of one mean, we simulated drawing a sample of specified size from
a specified g-and-h distribution with $h=0$, doing so 10,000 times
in each case to get an accurate estimate of proportion of Type I errors
in a two-tailed hypothesis test with alpha level 0.05. In each test,
we bootstrapped 1000 samples. The results are as follows:

\includegraphics[width=6cm,height=6cm]{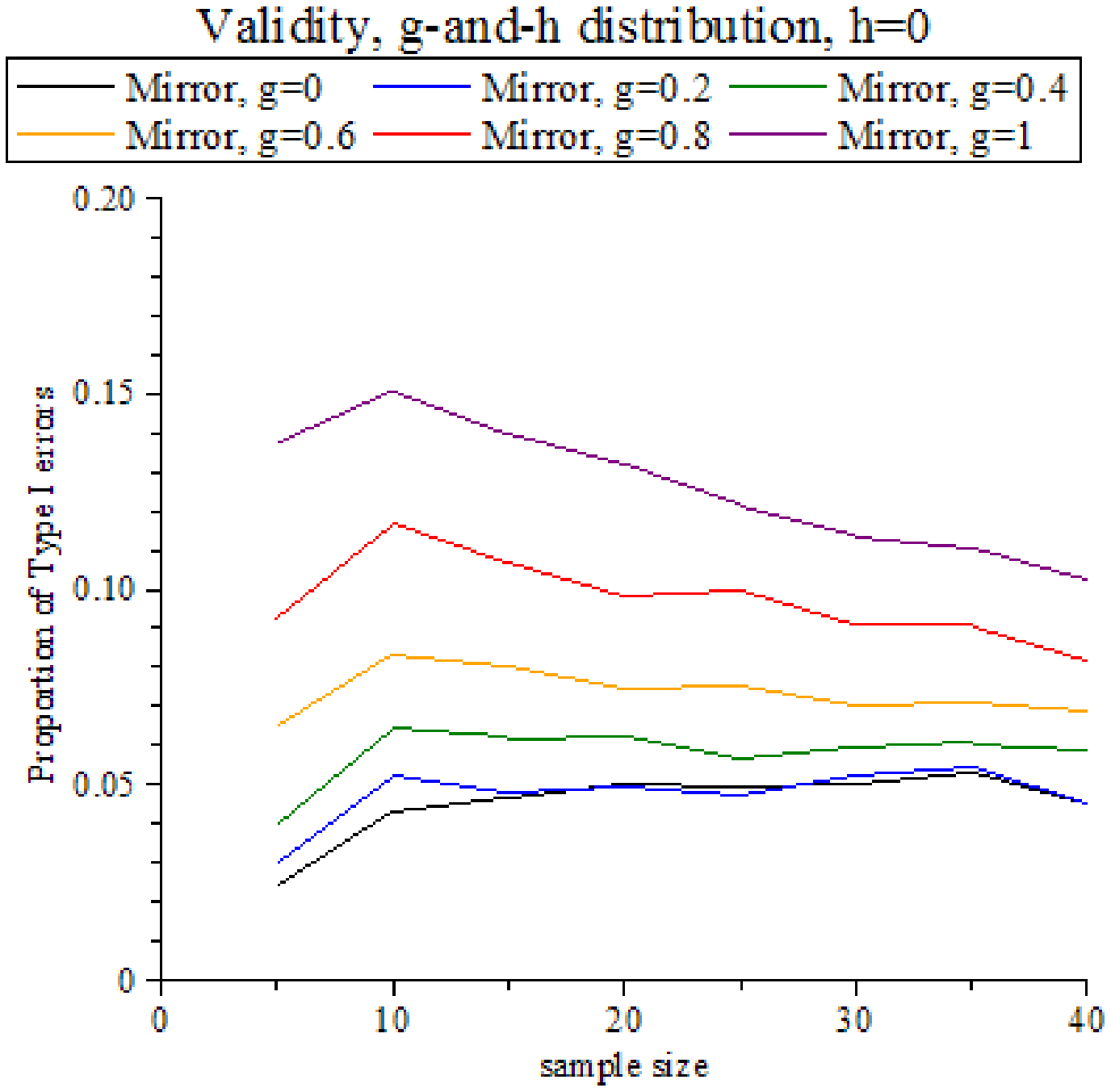}

As the skewness increases, the Mirror Bootstrap loses validity for
small samples, though the performance increases with sample size.
The validity is good for $g\le0.4$, and still decent for $g=0.6$,
but after that it becomes unacceptably high. Compared to the validity
of the t-test, Mirror Bootstrap performs as well as or better, especially
for small samples:

\includegraphics[width=6cm,height=6cm]{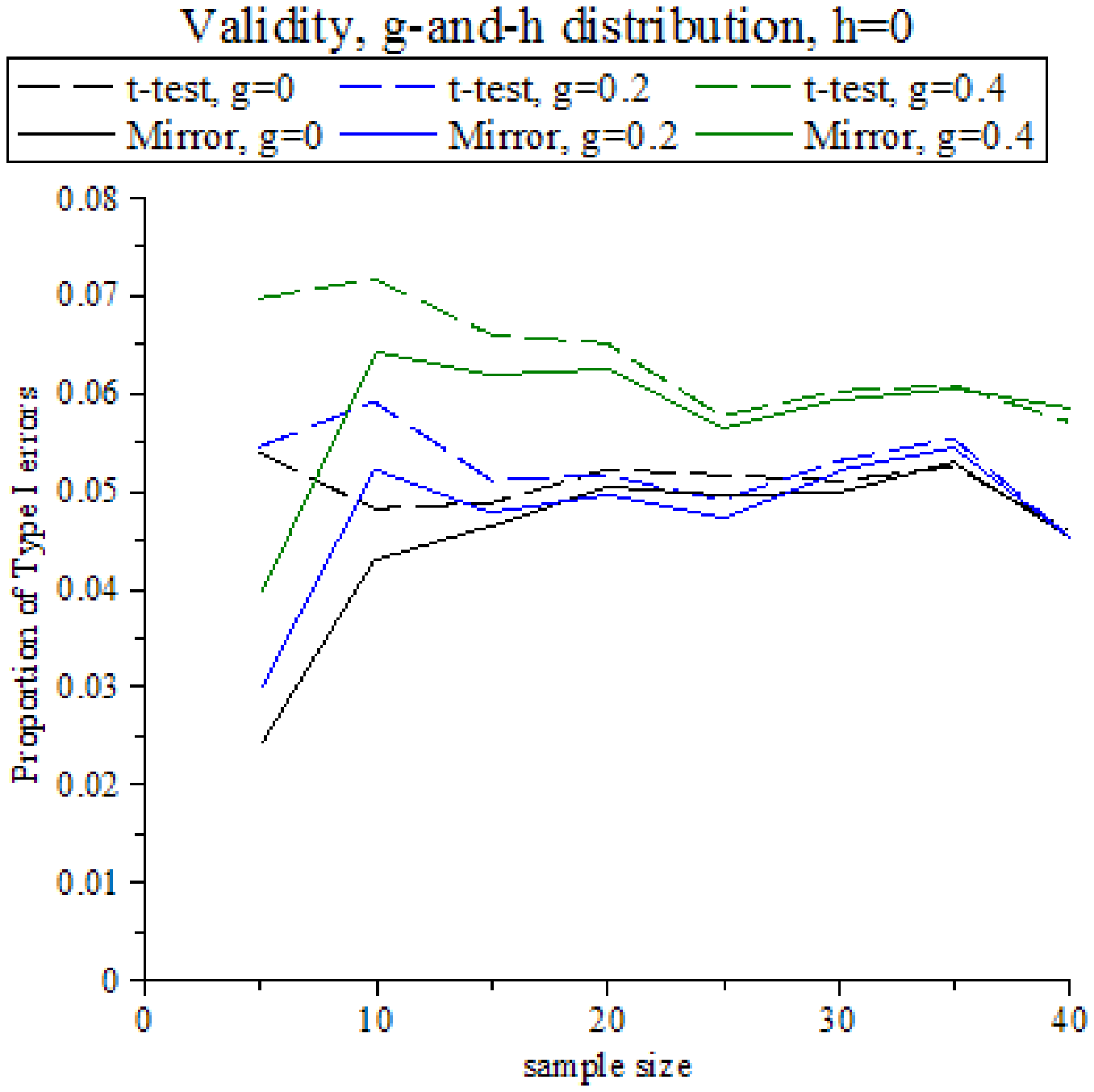}
\includegraphics[width=6cm,height=6cm]{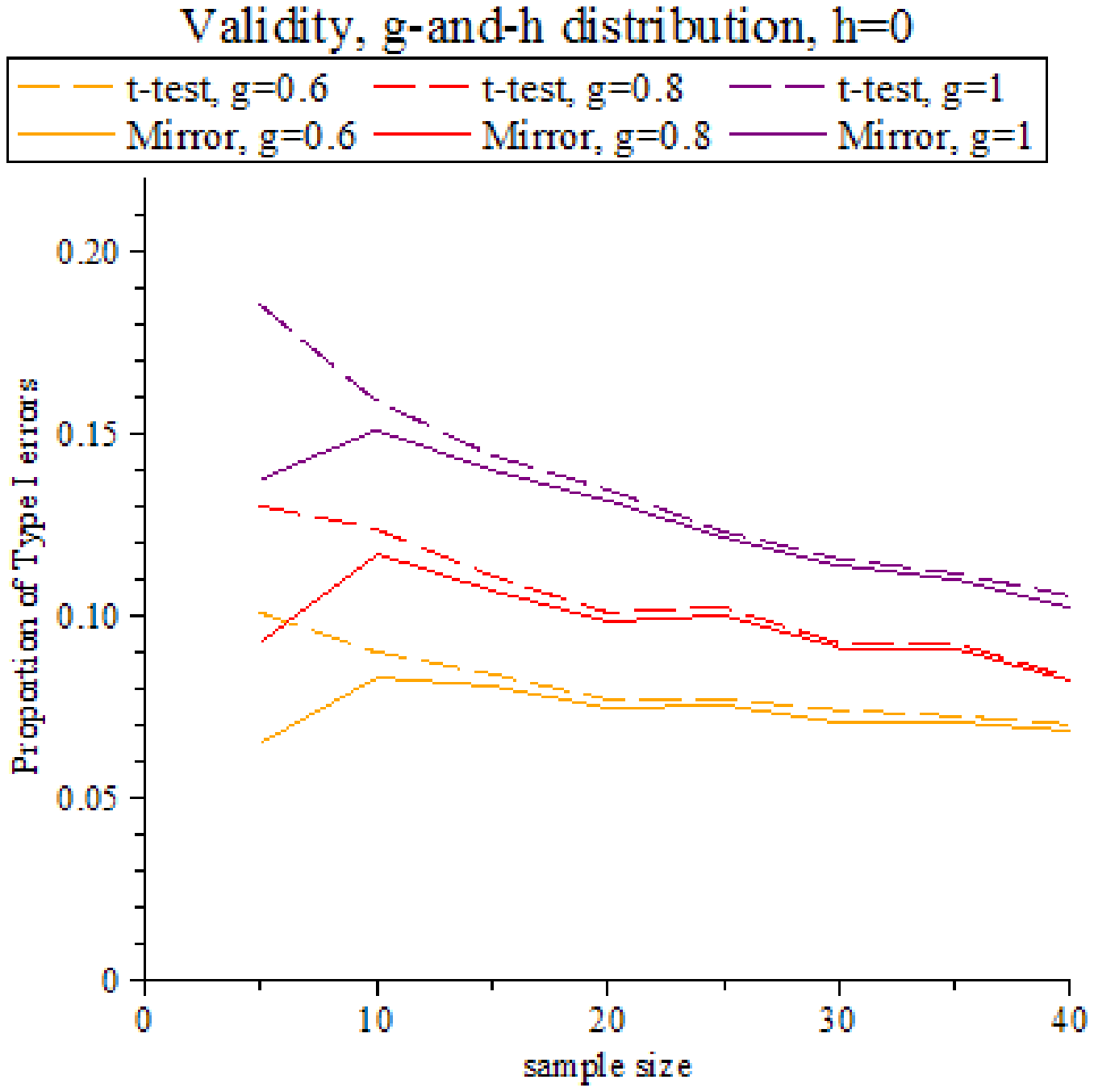}

\subsection{Heavy tails without skewness ($g=0$)}

To test the validity of the Mirror Bootstrap method in testing hypotheses
of one mean, we simulated drawing a sample of specified size from
a specified g-and-h distribution with $g=0$, doing so 10,000 times
in each case to get an accurate estimate of proportion of Type I errors
in a two-tailed hypothesis test with alpha level 0.05. In each test,
we bootstrapped 1000 samples. The results are as follows:

\includegraphics[width=6cm,height=6cm]{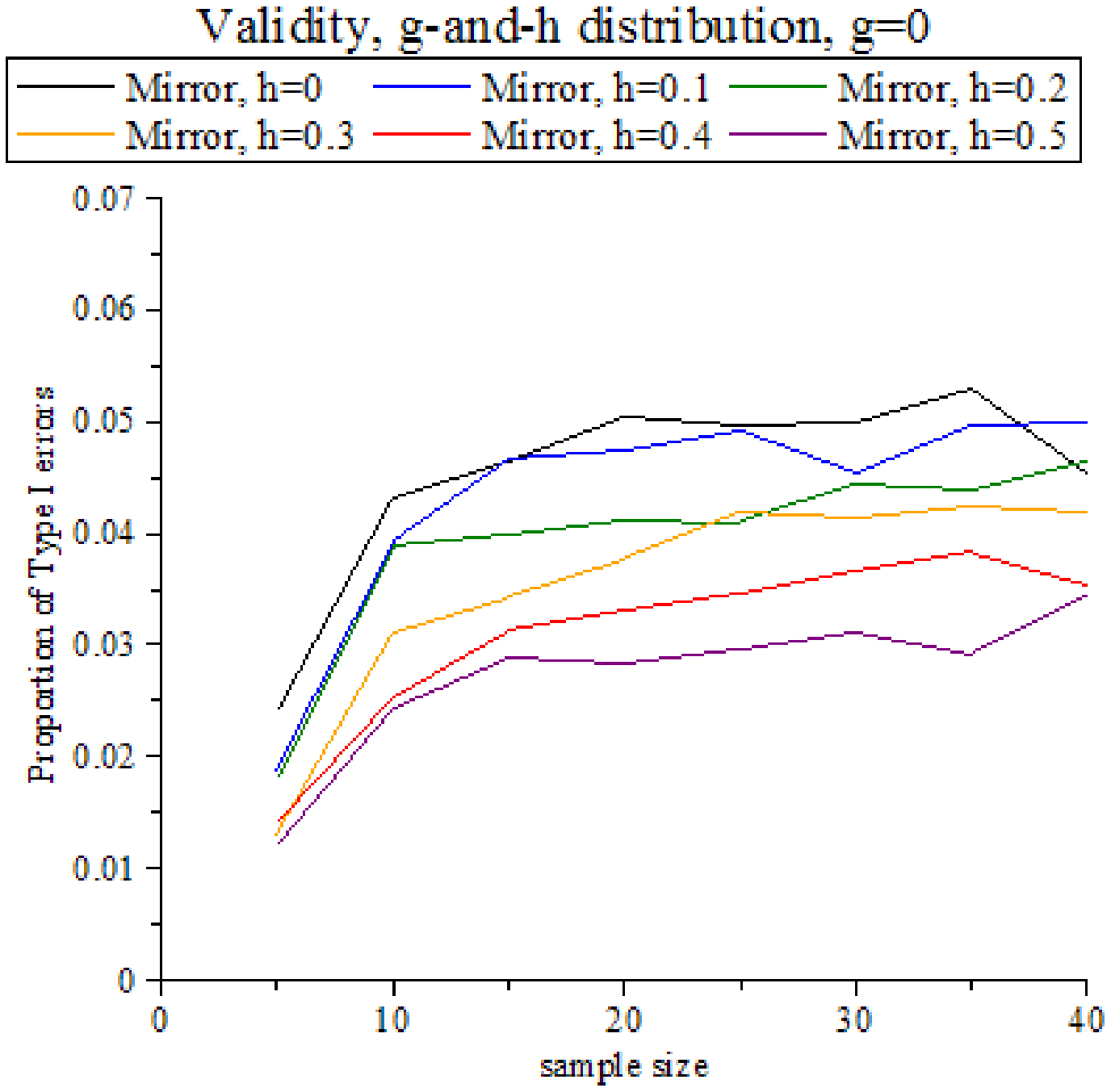}

Mirror Bootstrap method is conservative for very small samples ($n=5$),
and stays somewhat conservative for distributions with heavy tails
($h\ge4$), but for lighter tails the validity holds just below the
ideal level of 0.05, so the method is slightly conservative. Comparing
the Bootstrap Method's performance to that of the t-distribution,
the Mirror Bootstrap is consistently more conservative than the t-test,
though only slightly so for $n\ge10$:

\includegraphics[width=6cm,height=6cm]{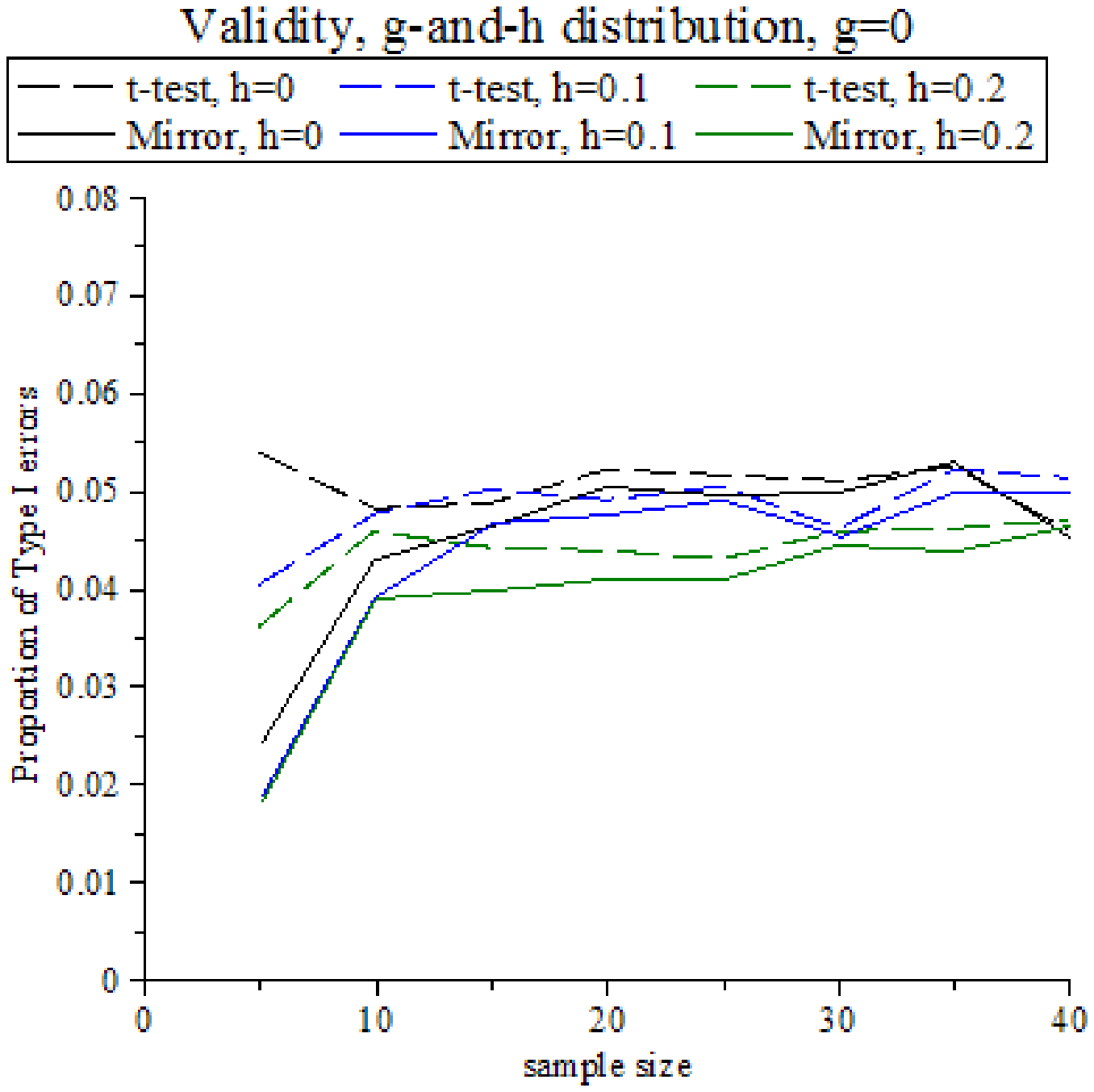}
\includegraphics[width=6cm,height=6cm]{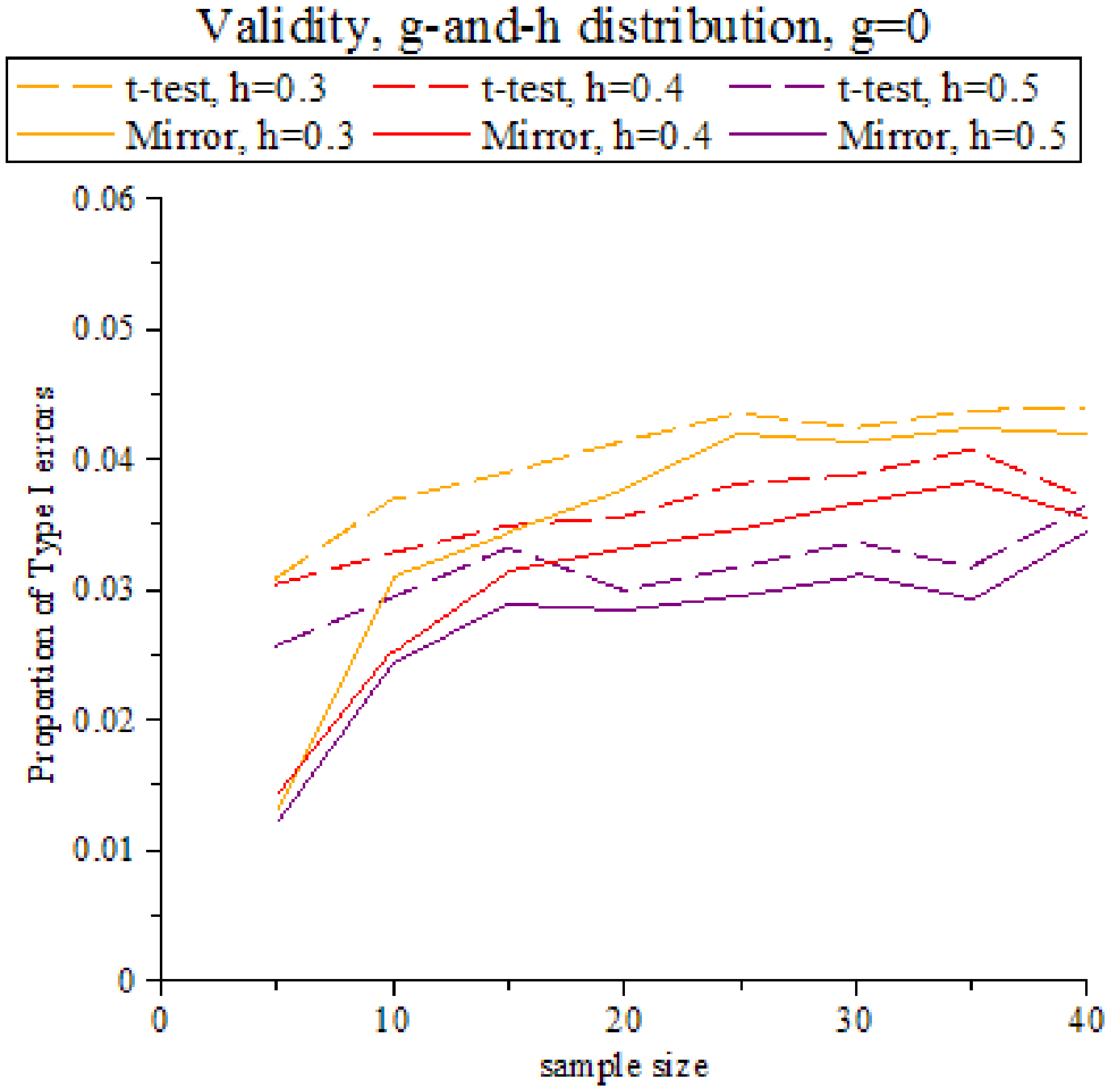}

Overall, the proportion of Type I errors appears reasonable for $g<0.6$
as long as the tails are not too heavy, and this result seems fairly
consistent for various sample sizes:

\includegraphics[width=6cm,height=6cm]{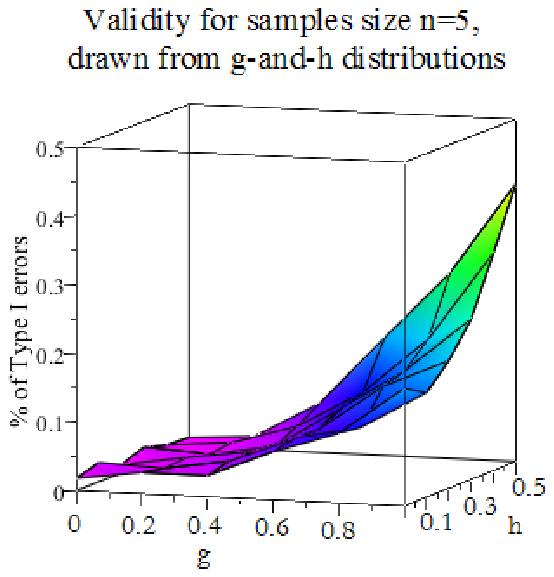}
\includegraphics[width=6cm,height=6cm]{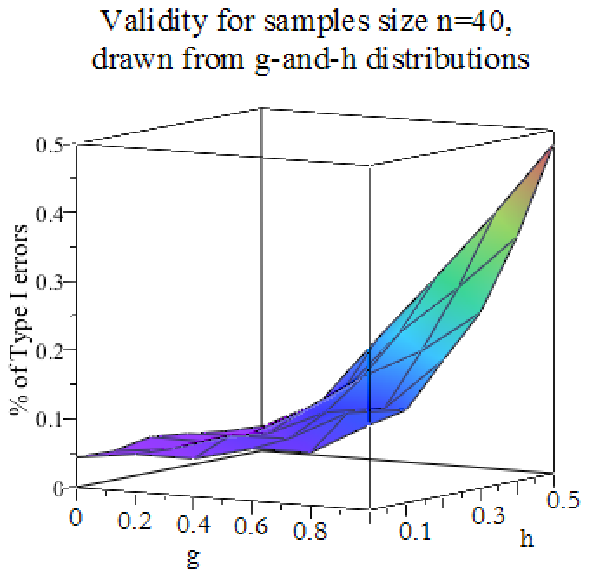}

\section{Acknowledgments}

The author would like to thank Dr. Wilcox for his helpful feedback,
and to the Lock5 team (Drs. Robin Lock, Patti Lock, Kari Lock, and
Masters Eric Lock and Dennis Lock, of www.lock5stat.com) for their
work in making bootstrapping methods readily accessible to teachers
of introductory statistics.

\end{document}